\documentclass[twocolumn,prl,aps,showpacs,floatfix,superscriptaddress]{revtex4}
\usepackage{graphicx}
\begin{document}

\widetext

\title{Divergence of the Chaotic Layer Width and Strong Acceleration
of the Spatial Chaotic Transport in Periodic Systems Driven by
an Adiabatic ac Force}

\author{S.M. Soskin}
\altaffiliation{Physics Department, Lancaster University,
UK}
\affiliation{Institute of Semiconductor Physics,
National Academy of Sciences of Ukraine, 03028 Kiev, Ukraine}
\affiliation{Abdus Salam ICTP, 34100 Trieste, Italy}
\author{O.M. Yevtushenko}
\affiliation{Abdus Salam ICTP, 34100 Trieste, Italy}
\author{R. Mannella}
\affiliation{Dipartimento di Fisica, Universit\`{a} di Pisa, 56127 Pisa, Italy}

\begin{abstract}
We show
for the first time that
a {\it weak} perturbation in a Hamiltonian system may
lead to an arbitrarily {\it wide} chaotic layer and {\it fast} chaotic transport.
This {\it generic} effect occurs in any spatially periodic Hamiltonian system subject
to
a
sufficiently slow ac force.
We explain it and develop an explicit theory
for the layer width,
verified in
simulations. Chaotic spatial transport as well as applications to the diffusion of
particles on surfaces, threshold devices and others are discussed.
\end{abstract}
\maketitle

The basic chaotic
formation
in perturbed Hamiltonian systems
is \cite{Zaslavsky:1991,lichtenberg_lieberman,zaslavsky:1998,E&E:book} a
chaotic layer associated with a separatrix of
the unperturbed (integrable) Hamiltonian system. Even in the
simplest case, when the unperturbed Hamiltonian system is
one-dimensional while the perturbation is time-periodic,
both the transport within the
layer \cite{E&E:book,vered,zaslavsky:2002} and
its structure on the
Poincar\'{e} section \cite{zaslavsky:1998,E&E:book,zaslavsky:2002}, relating to the homoclinic tangle, are very complicated. At
the same time, the boundaries of the layer are well defined
\cite{zaslavsky:1998} as the
last invariant curves which limit the layer from above and below
in the energy scale, and may be accurately
found in numerical simulations \cite{zaslavsky:1998}. For both
theory and applications, one of the
most important characteristics of the layer
is its width in
energy \cite{lichtenberg_lieberman,Zaslavsky:1991,zaslavsky:1998,soskin2000,PR,we}
or in related quantities \cite{E&E:book,E&E:1991,cary}.
It might be assumed that, if the perturbation is
very
{\it weak},
then the layer should necessarily be
{\it narrow}.
This natural assumption seems to be supported by numerous
examples (e.g.
\cite{Zaslavsky:1991, lichtenberg_lieberman,zaslavsky:1998,E&E:book,soskin2000,PR,we,E&E:1991,cary}).
However, we show
in the present Letter
that the situation may drastically differ (Fig. 1)
in a rather general case, namely for any {\it spatially periodic}
system \cite{footnote1} driven by a {\it slow ac
force}:
the upper energy boundary of the layer {\it diverges} as the frequency of
the force goes to zero.
We explain this,
develop a
theory
and verify it by
simulations. We also demonstrate that
the chaotic transport in space may be very
fast in the adiabatic case, on
sufficiently long time-scales,
and discuss
some
applications.

\begin{figure}[tb]
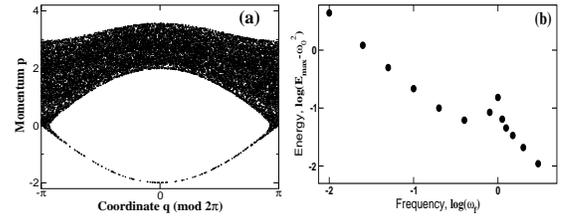

\includegraphics[width = 1.4 in]{Fig1a.eps}
   \includegraphics[width=1.4 in,height=1.1in]{Fig1b.eps}
\caption{
(a). The trajectory in the
stroboscopic (for $t=n2\pi/\omega_f$ with $n=0,1,2,...$)
Poincar\'{e} section for the system (1) with $\omega_0=1$, $h=0.01$, $\omega_f=0.01$
and initial conditions
$(p(0)=0,q(0)=\pi)$.
Number of points is 20000.
For the sake of compactness, we map all points onto the interval
$[-\pi,\pi[$: $q\rightarrow q-2\pi[(q+\pi)/(2\pi)]$. This mapping does not
affect the energy $ E\equiv p^2/2+U_0(q)$ and,
hence,
neither the chaotic layer width.
(b).
Spectral dependence of the  maximum $E_{max}$ of the
energy $E$ in the chaotic layer ($\omega_0=1,h=0.01$).
}
\end{figure}

Before moving on to the detailed consideration, we
comment on the relation
to
\cite{E&E:1991}.
It is shown in
\cite{E&E:1991} that, for a
system with adiabatically slowly pulsating parameters, the
homoclinic tangle covers most of the range swept by the
instantaneous separatrix.
If the pulsation of parameters in \cite{E&E:1991} were weak,
the range swept would be
{\it narrow}.
Our system essentially differs from that considered
in \cite{E&E:1991} (our perturbation
is not parametric) so the result of \cite{E&E:1991}
cannot be directly applied to it. If nevertheless
the result of \cite{E&E:1991} were
formally generalized to our system,
it would give that
the homoclinic tangle in the adiabatic limit covered the whole
phase space, thus hinting at a divergence of the
chaotic layer width.

Periodic systems and ac forces are widespread in nature.
An archetypal example \cite{Zaslavsky:1991}
is a pendulum driven by a
weak single-harmonic ac force:
\begin{eqnarray}
&&
{\dot q}= p, \quad\quad {\dot p}=-dU_0/dq-h\omega_0^2\sin (\omega_ft)\equiv -dU/dq, \nonumber \\
&& U_0\equiv U_0(q)=-\omega_0^2\cos(q), \\
&& U\equiv U(q,t)=U_0(q)+qh\omega_0^2\sin(\omega_ft),\quad\quad h\ll 1. \nonumber
\end{eqnarray}

To get some insight
into the physical origin of the phenomenon, and to
obtain a qualitative explanation of Fig. 1, we first consider the
{\it strong} adiabatic limit
\begin{equation}
\epsilon\equiv \omega_f/(h\omega_0)\ll 1,
\end{equation}
a much wider range of $\omega_f$
will be considered afterwards.

Given that
$\omega_f$ is small,
the system may be considered
as moving in the quasi-stationary
\lq\lq wash-board'' potential $U$ with slope
$h\omega_0^2\sin(\omega_ft)$. The
absolute value of the
slope
is $\sim h\omega_0^2$ for most of the time
while its sign is positive during odd half-periods of the perturbation,
changing to the opposite
one in the even half-periods.
If the system is initially at the top of
the potential barrier,
then,
owing to the
condition (2), even a small slope
$\sim h\omega_0^2$ is sufficient to
accelerate it
during the first half-period up to a negative velocity of {\it
large}
absolute value. In the second half-period, the
slope
changes its sign,
resulting in a braking effect,
so that
the velocity returns
close to zero
at the end of the period.
Assuming that the maximum
of the kinetic
energy ${\cal K}\equiv p^2/2$ greatly exceeds
the spatial modulation of $U(q,t)$,
\begin{equation}
{\cal K}_{max}\gg \omega_0^2,
\end{equation}
one may neglect the term
$U_0(q)$ in $U(q,t)$
while describing the major part of the trajectory (namely, when ${\cal K}\gg
\omega_0^2$) during the
first period of perturbation:
the equations of motion reduce to those of a free particle driven by
the
time-periodic force,
which are solved exactly:
\begin{eqnarray}
&&q(t)= A+
\left(B-\frac{h\omega_0^2}{\omega_f}\right)
t+\frac{h\omega_0^2}{\omega_f^2}\sin(\omega_ft),
\nonumber
\\
&&p(t)\equiv\dot{q}=B-\frac{h\omega_0^2}{\omega_f}(1-\cos(\omega_ft)),
\end{eqnarray}
where
$A=q(0)$ and $B=p(0)$.

For the given initial state
$(p(0)=0, q(0)=\pi)$, $B=0$ so that
the velocity (4) oscillates from 0 to $-2h\omega_0^2/\omega_f$
while the kinetic energy ${\cal K}\equiv p^2/2$ oscillates from 0 to
\begin{equation}
{\cal K}_{max}=2h^2\omega_0^4/\omega_f^2\equiv
2\omega_0^2/\epsilon^2,
\quad\quad \epsilon\ll 1.
\end{equation}

\noindent
Eq. (5) confirms the validity of the assumption (3), and
the maximum of the energy
does diverge as $\omega_f\rightarrow 0$.

But Eqs. (4),(5) do not explain chaos in the system.
Like in other cases \cite{Zaslavsky:1991,lichtenberg_lieberman,zaslavsky:1998,E&E:book},
the
chaos onset in our system
is related to the motion near the unperturbed
separatrix. Its rigorous treatment is
complicated (cf. the conventional adiabatic case \cite{E&E:1991}) but
is not essential for the quantity of main
interest in our paper, i.e. for the chaotic layer {\it width}:
the width is determined mainly by the
``regular'' parts of the chaotic trajectory (described
by (4)).
So, we describe the
chaos onset just qualitatively.



Define $h_n\equiv h_n(\omega_f)$ as the value of $h$ for
which the system
starting from the top of
the barrier ($p(0)=0, q(0)=\pi$)
arrives in the end of the first period of the perturbation
$t^{(1)}\equiv 2\pi/\omega_f$
at the top of the $n$th barrier i.e.
$\dot{q}(t^{(1)})=0$, $q(t^{(1)})=\pi-2\pi n$ where $n$ is
a large positive integer.
If $h=h_n$,
the velocity of the system at the
instant
when it passed the
previous (i.e. $(n-1)$st) barrier top
can be shown
to be
$-\dot{q}_c$ with $\dot{q}_c\sim \sqrt{h\omega_f\omega_0\ln
(\omega_0/(h\omega_f))
}
\ll\omega_0$.

Consider $h$
{\it slightly smaller} than $h_n$.
For any $t$,
$-\dot{q}(t)$
on the regular part of the trajectory
is slightly smaller than that for $h=h_n$. Hence,
the velocity on the true trajectory becomes zero slightly to the
right of the $n$th barrier
top slightly before $t^{(1)}$, i.e. the trajectory {\it reflects}
from the right slope of the
$n$th barrier and $\dot{q}(t^{(1)})>0$. The new round of
acceleration to the left starts only after the next reflection
(from the left slope of the $(n-1)$st barrier) i.e. with a small
delay $\Delta t_1$ with respect to $t^{(1)}$:

\begin{equation}
\Delta t_1\sim\omega_0^{-1}\ln(\omega_0^2/\dot{q}_c^2)\sim\omega_0^{-1}\ln(\omega_0/(h\omega_f))\ll
\omega_f^{-1}.
\end{equation}

\noindent
This new
round of acceleration-braking
is described by Eq. (4)
with $B$ determined by the condition
$p(2\pi/\omega_f+\Delta t_1)=0$, i.e.
$B=
(\omega_0/\epsilon)
(1-\cos(\omega_f\Delta t_1))>0$.

Consider $h$
{\it slightly larger} than $h_n$. For any $t$, $-\dot{q}(t)$
is slightly
larger than that for $h=h_n$  on the regular part of the trajectory. Hence, at
$t^{(1)}$,
the system
arrives
slightly to the left of the top of the
$n$th barrier while moving to the left and with an
energy
close to the top of the barrier level.
When the coordinate of the system approaches
the vicinity of the top of the next (i.e. $(n+1)$st)
barrier of $U(q,t)$, the latter becomes sufficiently lower than the top of
the $n$th barrier and the system passes over it rather than reflects. Thus, the
second round of acceleration-braking is described by Eq. (4)
with $B\equiv \dot{q}(t^{(1)})$ being negative with a small absolute value.

At the end of the second acceleration-braking round,
the system may again either pass over a barrier or
reflect from it and even pass in the backward
direction for a few periods of the potential. In the latter case,
the new acceleration-braking round is additionally
delayed with respect to the perturbation
by some time $\Delta t_2$:
this round is described by (4)
with yet a larger positive
$B=(\omega_0/\epsilon)(1-\cos(\omega_f[\Delta t_1+\Delta t_2]))>0$.
As time goes on,
the process of
reflection develops and, on long time-scales, the delay due
to the reflections is {\it accumulated}.
So, $B$
gradually growes in average until it
gets close to $B_{max}\equiv
2\omega_0/\epsilon$, then gradually
decreases in average until it gets close to
0,
etc.
The sequence of reflections/passings is
random unless $h=h_n$, but even in this case the sequence
is random if the initial state is shifted from the
top of the barrier.
As reflections/passings
and
intervals $\Delta t_i$
are random,
variations of $B$ are random too:
so,
the trajectory is {\it chaotic}
on large time-scales.

Note also that the action $I\equiv (2\pi)^{-1}\oint p \;{\rm d} q $ \cite{cyclic},
conventionally \cite{E&E:book,E&E:1991,cary} chosen as
the lowest-order adiabatic invariant,
is {\it not conserved} for motion above the barrier in our
system: on the major part of a trajectory, $K\gg \omega_0^2$ and
hence $I\approx p$ while $|p|$ varies in a {\it wide} range
(from 0 to $\sim \omega_0/\epsilon$, for the chaotic trajectory).
The correct lowest-order {\it adiabatic invariant} for our system is

\begin{equation}
\tilde{I}=I+\frac{h\omega_0^2}{\omega_f}(1-\cos(\omega_ft)).
\end{equation}

\noindent
To the lowest order in $\epsilon$, $\tilde{I}$
coincides with the integration constant $B$ of (4) on the major part of the trajectory \cite{separatrix_crossing}. So, the chaotic layer
width in
$\tilde{I}$ is the same as in $B$, i.e. equal
to $B_{max}\equiv 2h\omega_0^2/\omega_f$,
{\it diverging} as $\omega_f\rightarrow 0$.

Let us consider the problem of the layer width in a
more general case, when
Eq.
(2)
may not be
satisfied.
As
explained
above, the trajectory
that can both pass over a barrier and reflect from it
is chaotic.
But
the strong
inequality (3) for ${\cal K}_{max}$ may
not hold now (cf. Eq. (5)) so
the term $U_0$ in $U$
cannot be neglected even on
regular
parts of
the chaotic trajectory.
Let the adiabatic approximation be still valid
(the explicit condition is derived in
Eq. (18)).
As
$h$ is small, the turning point of the chaotic trajectory
is situated
near the
{\it top} of a potential barrier:

\begin{equation}
p(t_s)=0,\quad\quad q(t_s)\approx \pi-2\pi m,
\end{equation}

\noindent
where $m$ is integer
and $t_s$ is one of instants when $p=0$.

Let ${\cal K}_n$ be the kinetic energy ${\cal K}$ at the
instant $t_n$ when the
system crosses the coordinate of the top of the $n$th barrier of the
auxiliary
potential $U$, i.e.
\begin{equation}
q(t_n)\equiv q_n\approx \pi-2\pi n,
\quad\quad
{\cal K}_n\equiv \frac{p^2(t_n)}{2}.
\end{equation}
The coordinate of the next barrier
top crossed by the system is

\begin{equation}
q_{n+\delta_n}\approx q_n-2\pi\delta_n,
\quad\quad
\delta_n\equiv -{\rm sign}(p(t_n)).
\end{equation}
In the adiabatic approximation, the change
of ${\cal K}$ during the time $t_{n+\delta_n}-t_n$ is the following:
\begin{equation}
{\cal K}_{n+\delta_n}-{\cal K}_n\approx
-(q_{n+\delta_n}-q_n)h\omega_0^2\sin(\omega_ft_n).
\end{equation}
The interval of time $t_{n+\delta_n}-t_n$ can be evaluated
in the adiabatic approximation as
follows
\begin{eqnarray}
&&t_{n+\delta_n}-t_n\equiv\int_{q_n}^{q_{n+\delta_n}}\frac{{\rm
d}q}{p} \\
&&\approx
\int^{q_n}_{q_{n+\delta_n}}\frac{{\rm d}
q}{\delta_n
\sqrt{2({\cal K}_n+\omega_0^2(1+\cos(q)))}}\approx
\frac{2\sqrt{x} {\rm K}(x)}{\omega_0},
\nonumber \\
&&{\rm K}(x)
\equiv
\int_{0}^{\frac{\pi}{2}}\frac{d\varphi}{\sqrt{1-x\sin^2(\varphi)}}, \quad\quad
x\equiv \frac{1}{1+
{\cal K}_n/(2\omega_0^2)
}.
\nonumber
\end{eqnarray}
${\rm K}(x)$ is the full elliptic integral of the 1st
order \cite{Abramovitz_Stegun}.

Significant changes of ${\cal K}_n$ take place on the
time scale $\omega_f^{-1}$ while the adiabatic condition means that
\begin{equation}
t_{n+\delta_n}-t_n\ll\omega_f^{-1}.
\end{equation}
So, {\it discrete} changes of ${\cal K}_n \equiv {\cal K}(t_n)$
may be replaced by the {\it continuous}
change of a function $\tilde{\cal K}(t_n)$ with derivative
\begin{eqnarray}
&&
\frac{{\rm d}{\tilde{\cal K}}}{{\rm d}t_n}
\approx
\frac{{\cal K}_{n+\delta_n}-{\cal K}_n}
{t_{n+\delta_n}-t_n} \nonumber\\
&&
\approx -{\rm sign}(p(t_n))
\frac{\pi
h\omega_0^3\sqrt{1+\tilde{\cal K}/(2\omega_0^2)}}{{\rm K}(1/(1+\tilde{\cal K}/(2\omega_0^2)))}
\sin(\omega_ft_n).\quad
\end{eqnarray}
Separating variables,
evaluating the resulting integrals
and
taking into account that, within the present simplified
description of motion, ${\rm sign}(p)$ changes when $\tilde{\cal K}=0$,
we obtain the
transcendental equation for $\tilde{\cal K}$:

\begin{eqnarray}
&&\frac{{\rm E}(x)}{\sqrt{x}}=1+\frac{\pi h}{4}\frac{\omega_0}{\omega_f}|\cos(\omega_ft_s)-\cos(\omega_ft_n)|,\\
&&{\rm E}(x)\equiv \int_0^{\frac{\pi}{2}}{\rm d}\varphi\;\sqrt{1-x\sin^2(\varphi)},\quad
x\equiv \left(1+\frac{\tilde{\cal K}}{2\omega_0^2}\right)^{-1}.
\nonumber
\end{eqnarray}
Here,
${\rm E}(x)$ is a full elliptic integral of the 2nd
order \cite{Abramovitz_Stegun}.

Due to the random-like changes of $t_s$,
$\cos(\omega_ft_s)$ on long time-scales is densely
distributed over the range $[-1,1]$.
Maximizing $\tilde{\cal K}$
with respect to $t_s$ and $t_n$ and taking into account that
$E(t_n)\equiv {\cal K}(t_n)+U_0(q(t_n))\approx \tilde{\cal
K}(t_n)+\omega_0^2,$
we finally obtain
the transcendental equation for the
upper energy boundary of the chaotic layer $E_{max}$:
\begin{equation}
\frac{{\rm E}(x)}{\sqrt{x}}=1+\frac{\pi h}{2}\frac{\omega_0}{\omega_f},
\quad\quad
x\equiv \left(1+\frac{E_{max}-\omega_0^2}{2\omega_0^2}\right)^{-1}.
\end{equation}
$E_{max}$
monotonously
decreases from
$\infty$
to $\omega_0^2$  as
$\omega_f/h\omega_0\equiv \epsilon$
increases from 0 to
$\infty$.
Fig. 2 shows that
Eq. (16) nicely describes the
simulations in a wide range of $\omega_f$.

\begin{figure}[htb]
   \begin{center}
   \includegraphics[width=4.cm,height=2.5cm]{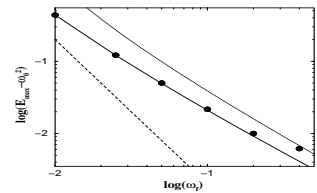}
   \end{center}
      \caption[fig5]
   {\label{fig:fig5}
   Spectral dependence of the maximum energy in the chaotic
layer ($\omega_0=1, h=0.01$):
circles, solid, dashed and dotted lines correspond respectively to simulations, the general formula (16),
the asymptotes (5)
and (17).
}
\end{figure}

For
$\epsilon\ll 1$,
the root $(E_{max}-\omega_0^2)/\omega_0^2$ of Eq. (16) is large.
So,
$x\approx 2\omega_0^2/E_{max}\rightarrow 0$
while ${\rm E}(x\rightarrow 0)\rightarrow \pi/2$, and the
solution of (16) reduces to the asymptote (5).

For
$\epsilon\gg 1$,
the root $(E_{max}-\omega_0^2)/\omega_0^2$ of Eq. (16) is small.
Using the asymptote \cite{Abramovitz_Stegun} for ${\rm E}(x\rightarrow
1)$, Eq. (16) can be reduced to the
following asymptote
for $E_{max}$:
\begin{equation}
\frac{E_{max}-\omega_0^2}{\omega_0^2}\approx \frac{32}{a\ln(a)},
\quad a\equiv \frac{8\epsilon}{\pi}, \quad \epsilon\gg 1.
\end{equation}

Using the
estimate (12) and the asymptote
${\rm K}(x \rightarrow 1) = 0.5\ln(16/(1-x))$ \cite{Abramovitz_Stegun}
and allowing for Eq. (17),
we may express the adiabatic condition (13) explicitly:
\begin{equation}
\omega_f\ll
\omega_0/\ln(1/h).
\end{equation}

Let us discuss
chaotic transport.
For most physical applications, transport in coordinate is relevant.
For sufficiently long time-scales,
the chaotic trajectory provides
large-scale displacements
in {\it both} directions,
unlike regular trajectories.
Generally,
the mean-square displacement
for chaotic transport
depends on time as
\cite{zaslavsky:2002}
$\langle(q(t)-q(0))^2\rangle=Dt^{b}$ with
$0<b<2$
($\langle ...\rangle$ means averaging of the initial
conditions over the chaotic layer).
The {\it larger}
$D$ and $b$ are, the {\it faster} the
transport
is.
It is suggested in \cite{india}, basing on numerical results,
that $b\rightarrow 1$
as $\omega_f\rightarrow 0$.
As for $D$, we suggest
that it {\it strongly diverges}.
Indeed,
chaotic trajectories generally spend most of the
time close to the boundaries of
regions of regular motion \cite{zaslavsky:2002}. In our case, it
means that, for most of the
time, the trajectory moves
close to either the
upper border of the chaotic layer, with the average velocity
$v\approx h\omega_0^2/\omega_f$,
or the lower border, with the average velocity $-v$.
We call such regimes
{\it acceleration-braking flights}
(cf. L\'{e}vy flights \cite{zaslavsky:2002,shlesinger}),
distinguishing them from the regime of
diffusion across the layer.
The duration $t_{f}$ of the flight may be estimated from the
analysis of the diffusion of $B$
near
the boundary of the layer. The diffusion constant
for $B$ may be roughly estimated as
$D_B\sim\langle (\Delta B)^2 \rangle/(2\pi/\omega_f)$ where
$\langle (\Delta B)^2 \rangle $ is the average squared change
of $B$
at the end of a driving period:
$\langle (\Delta B)^2 \rangle \sim B_{max}^2(\omega_f\Delta
t_1)^4$.
Then,
\begin{equation}
t_{f}\sim \frac{B_{max}^2}{D_B}\sim\omega_0^{-1}\frac{(\omega_0/\omega_f)^5}{\ln^4(\omega_0/(h\omega_f))}.
\end{equation}

\noindent
Finally, $D$ may be estimated as the ratio between
the squared length of the flight $l_{f}^2$ and its duration $t_{f}$:
\begin{equation}
D\sim \frac{ l_{f}^2}{t_{f}}\equiv v^2t_{f}\sim \omega_0h^2\frac{(\omega_0/\omega_f)^7}{\ln^4(\omega_0/(h\omega_f))}.
\end{equation}

\noindent
The above analysis provides intuitive arguments in favor of a strong acceleration
of the chaotic transport in space as $\omega_f\rightarrow 0$,
and simulations
support this.
Still, a thorough
numerical study and
a rigorous
evaluation
of $D(\omega_f)$
as well as a proof that
$b(\omega_f\rightarrow 0)=1$
are necessary.

Apart from purely dynamic phenomena,
our work may have a strong impact on noise-induced phenomena: e.g.
diffusion of particles on surfaces
(see \cite{katya} and
references therein),
that
plays an important role
in many modern technologies involving self-assembled molecular-film
growth, catalysis, and surface-bound nanostructures \cite{Science_etc}.
Numerous studies of atoms (e.g. \cite {diffusion_atoms}), organic molecules (e.g. \cite
{diffusion_organic_molecules}), and even metal clusters
(e.g. \cite{diffusion_clusters}) show that the {\it long} jumps may
play a dominant role in
the
diffusion: this
means that the {\it damping is small}.
If we add a weak ac drive, then transient chaos
\cite{lichtenberg_lieberman} arises in a region of phase space which
approximately coincides with the chaotic layer developed for zero damping
(cf. \cite{soskin2000}). In the light of the present results, the latter
means: if the driving is {\it adiabatic},
then, as
soon as the noise activates the particle to the lower boundary of the
layer (i.e. approximately to the barrier level), the further transport
is provided by fast (compared to slow noise-induced diffusion)
chaotic transport. Thus, even a {\it weak}
drive may drastically
increase the speed of the diffusion.

One more
possible application concerns
a {\it threshold} device (cf. \cite{threshold_devices}). If the device was
based on a noise-driven spatially periodic system and switched at the energy
level $E_{th}$ significantly exceeding the barrier level $U_b$, then the
addition of even a weak but sufficiently slow ac drive would give rise to a
marked decrease
of the activation energy - from $E_{th}$ to $U_b$, thus leading to a drastic
increase of the flux. This may also be used as a
sensitive method
to measure
the amplitude/frequency of the drive.


In conclusion, we have discovered the
divergence of the chaotic layer width and the related strong acceleration of
the chaotic spatial transport in ac driven periodic systems in the adiabatic limit.
The mechanism is a combination of acceleration-braking flights of
high average kinetic energy
and of small random changes of the adiabatic invariant
near
the separatrix.
Applications to diffusion on surface and to threshold devices
have been
suggested.

\end{document}